\documentclass[preprintnumbers,amsmath,amssymb,nofootinbib]{revtex4}

\usepackage{graphicx}
\usepackage{epsfig}
\usepackage{color}

\newcommand{\beq}{\begin{equation}}
\newcommand{\eeq}{\end{equation}}
\newcommand{\bea}{\begin{eqnarray}}
\newcommand{\eea}{\end{eqnarray}}
\newcommand{\bwd}{\begin{widetext}}
\newcommand{\ewd}{\end{widetext}}

\begin{document}

\title{Generalized space-charge solver for three-dimensional converging or diverging beams}


\author{Ji Qiang}
\email{jqiang@lbl.gov}
\affiliation{Lawrence Berkeley National Laboratory, Berkeley, CA 94720, USA}
%
\begin{abstract}

Nonlinear space-charge effects play a critical role in the dynamics of high-intensity, high-brightness beams. While conventional space-charge forces scale inversely with the square of the relativistic Lorentz factor ($\gamma$) as beam energy increases, this suppression may not apply to all components of the electromagnetic force. In this paper, we present a generalized space-charge solver that accounts for contributions from transverse currents in three-dimensional (3D) converging or diverging beams. We demonstrate that these additional forces from transverse currents lack an explicit dependence on $\gamma$, potentially rendering them significant in specific high-energy applications. We provide a semi-analytical solution for a 3D converging or diverging beam with the Gaussian density distribution and identify the regimes in which these transverse current forces become comparable to conventional space-charge forces. Furthermore, we present a numerical solution for the generalized model applicable to arbitrary density distributions. Finally, we evaluate the impact of these forces on strongly converging electron beams within the interaction regions of the FCC-ee and ILC, as well as the second bunch compressor of LCLS-II-HE; our results indicate that these extra forces do not significantly alter the final electron beam quality in these specific cases.

\end{abstract}

\maketitle

\section{Introduction}

Nonlinear space-charge effects arising from Coulomb interactions among charged particles can significantly constrain high-intensity and high-brightness accelerators. These effects may drive emittance growth, induce halo formation, and—under unfavorable conditions—lead to particle loss along the machine.

In the conventional space-charge model, the space-charge forces are suppressed at high energy and scale as $1/\gamma^{2}$, where $\gamma$ is the relativistic Lorentz factor~\cite{parmila,friedman,machida,pichoff,parmela,toutatis,galambos,franchetti03,impact,impact-t,amundson,opal,ferrario,adrian22}. Because the beam travels relativistically along the longitudinal direction, it is common to assume that the beam distribution is frozen in the moving beam frame so that the interaction reduces to static Coulomb forces. The space-charge force can be expressed in terms of the electrostatic potential together with the longitudinal component of the vector potential for a moving beam in the laboratory frame. The resulting interplay yields a cancellation of the purely electrostatic contribution and an overall $1/\gamma^{2}$ dependence as the beam energy increases.

The conventional description, however, may become incomplete when the beam undergoes strong transverse convergence. In particular, when a beam is focused from a large transverse size to a small one within the interaction region of colliders such as FCC-ee~\cite{fccee} or the ILC~\cite{ilc}, or when it converges to a small transverse size upstream of the final bending magnet in a chicane bunch compressor (e.g., after dispersion removal)~\cite{tor,lcls2he}, the transverse current may become non-negligible. Transverse currents generate additional electromagnetic forces through the associated magnetic fields, which are not captured by the standard space-charge model. To include these effects in high brightness beams, one must solve the wave equation for the electric potential given the charge density, together with the wave equations for the vector potentials given the current distribution.

Space-charge effects in converging or diverging relativistic beams have been studied using two-dimensional coasting-beam models~\cite{frank,bane}. In this work, we develop and present a generalized space-charge solver that accounts for transverse currents in a three-dimensional converging or diverging bunched beam.

The paper is organized as follows: Section II discusses the generalized space-charge model for 3D bunched beams. Section III presents the solver for a 3D Gaussian density distribution, and Section IV covers arbitrary density distributions in a straight accelerator system. Section V provides three applications of the solver to highly converging beams, and Section VI offers our conclusions.

\section{Generalized space-charge model for 3D converging/diverging beams}
We begin with the three-dimensional Maxwell equations for the electric potential and the magnetic vector potential to account for transverse currents in a three-dimensional converging or diverging beam. Using the Lorenz gauge, these equations can be written as:
\begin{eqnarray}
\frac{\partial^2 \phi}{\partial x^2} +
\frac{\partial^2 \phi}{\partial y^2} + \frac{\partial^2 \phi}{\partial s^2} - \frac{1}{c^2}\frac{\partial^2 \phi}{\partial t^2} & = &- \frac{\rho(x,y,s,t)}{\epsilon_0}  \\
\frac{\partial^2 A_x}{\partial x^2} +
\frac{\partial^2 A_x}{\partial y^2} + \frac{\partial^2 A_x}{\partial s^2} - \frac{1}{c^2}\frac{\partial^2 A_x}{\partial t^2} &= &- \frac{J_x(x,y,s,t)}{c^2\epsilon_0}  \\
\frac{\partial^2 A_y}{\partial x^2} +
\frac{\partial^2 A_y}{\partial y^2} + \frac{\partial^2 A_y}{\partial s^2} - \frac{1}{c^2}\frac{\partial^2 A_x}{\partial t^2} &=& - \frac{J_y(x,y,s,t)}{c^2\epsilon_0}  \\
\frac{\partial^2 A_s}{\partial x^2} +
\frac{\partial^2 A_s}{\partial y^2} + \frac{\partial^2 A_s}{\partial s^2} - \frac{1}{c^2}\frac{\partial^2 A_x}{\partial t^2} &=& - \frac{J_s(x,y,s,t)}{c^2\epsilon_0} 
\end{eqnarray}
where x and y denote the transverse coordinates, s denotes the longitudinal coordinate, $\phi$ is the electric scalar potential, $A_{x,y,s}$ are the three components of the magnetic vector potential. Here, c is the speed of light in vacuum, $\epsilon_0$ is the
vacuum permittivity, $\rho$ is the charge density, and $J_{x,y,s}$ are the three
components of the current density.
Assuming that 
the particles move predominantly along the s with a large longitudinal speed $v_0$, we
write 
\[
\rho=\rho(x,y,s-v_0 t), 
\qquad
\mathbf{J}=\mathbf{J}(x,y,s-v_0 t).
\]
Introducing a new variable
$z =s - v_0t$, Maxwell's equations can then be rewritten as:
\begin{eqnarray}
\label{ep1}
\frac{\partial^2 \phi}{\partial x^2} +
\frac{\partial^2 \phi}{\partial y^2} + \frac{1}{\gamma^2}\frac{\partial^2 \phi}{\partial z^2}  &=& - \frac{\rho(x,y,z)}{\epsilon_0}  \\
\frac{\partial^2 A_x}{\partial x^2} +
\frac{\partial^2 A_x}{\partial y^2} + \frac{1}{\gamma^2}\frac{\partial^2 A_x}{\partial z^2}  &=& - \frac{J_x(x,y,z)}{c^2\epsilon_0}  \\
\frac{\partial^2 A_y}{\partial x^2} +
\frac{\partial^2 A_y}{\partial y^2} + \frac{1}{\gamma^2}\frac{\partial^2 A_y}{\partial z^2}  &=& - \frac{J_y(x,y,z)}{c^2\epsilon_0}  \\
\frac{\partial^2 A_s}{\partial x^2} +
\frac{\partial^2 A_s}{\partial y^2} + \frac{1}{\gamma^2}\frac{\partial^2 A_s}{\partial z^2}  &=& - \frac{J_s(x,y,z)}{c^2\epsilon_0}  
\label{ep4}
\end{eqnarray}
where $\gamma=1/\sqrt{1-(v_0/c)^2}$ is the relativistic factor, 
and we have used $\frac{\partial}{\partial s} = \frac{\partial}{\partial z}$ and $\frac{\partial}{\partial t} = -v_0 \frac{\partial}{\partial z}$ in the above equation.
Further Assuming that there is no longitudinal relative motion
in the moving frame with a speed $v_0$, then the longitudinal current density
is $J_s = \rho v_0$, and the longitudinal component of the vector potential can be written as:
\begin{equation}
    A_s(x,y,z) = \frac{\beta}{c}\phi(x,y,z)
    \label{As}
\end{equation}
where $\beta=v_0/c$. 
Under the Lorenz gauge, the electromagnetic fields can be obtained from electric potential
and vector potential as:
\begin{eqnarray}
   \label{EMe}
   \bf{E} & = &  -\nabla \phi - \frac{\partial \bf{A}}{\partial t}  \\
   \bf{B} & = & \nabla \times \bf{A}
   \label{EMb}
\end{eqnarray}
From the electromagnetic fields above, the Lorentz forces $\bf{F}$ acting on an individual particle of charge $q$ are:
\begin{equation}
    {\bf{F}} = q (\bf{E}+v\times B)
\end{equation}
Substituting Eqs.~~\ref{EMe} and \ref{EMb} into the above equation, the Lorentz force can be written component-wise as:
\begin{eqnarray}
F_x & = & q(-\frac{\partial \phi}{\partial x}+v_0\frac{\partial A_x}{\partial z}+
    v_y\frac{\partial A_y}{\partial x}-v_y\frac{\partial A_x}{\partial y} - v_s\frac{\partial A_x}{\partial s} + v_s\frac{\partial A_s}{\partial x}) \\
F_y & = & q(-\frac{\partial \phi}{\partial y}+v_0\frac{\partial A_y}{\partial z}+
    v_s\frac{\partial A_s}{\partial y}-v_s\frac{\partial A_y}{\partial s} - v_x\frac{\partial A_y}{\partial x} + v_x\frac{\partial A_x}{\partial y}) \\  
F_s & = & q(-\frac{\partial \phi}{\partial s}+v_0\frac{\partial A_s}{\partial z}+
    v_x\frac{\partial A_x}{\partial s}-v_x\frac{\partial A_s}{\partial x} - v_y\frac{\partial A_s}{\partial y} + v_y\frac{\partial A_y}{\partial s})   
\end{eqnarray}
where $v_{x,y,s}$ are the transverse and longitudinal components
of the particle velocity.
Making use of the equation~\ref{As}, together with $\frac{\partial}{\partial s} = \frac{\partial}{\partial z}$ and assuming $v_s = v_0$,
the above equations can be reduced to:
\begin{eqnarray}
F_x & = & q(-\frac{1}{\gamma^2}\frac{\partial \phi}{\partial x}+
    v_y\frac{\partial A_y}{\partial x}-v_y\frac{\partial A_x}{\partial y})  \\
F_y & = & q(-\frac{1}{\gamma^2}\frac{\partial \phi}{\partial y}- v_x\frac{\partial A_y}{\partial x} + v_x\frac{\partial A_x}{\partial y}) \\  
F_s & = & q(-\frac{1}{\gamma^2}\frac{\partial \phi}{\partial z}+
    v_x\frac{\partial A_x}{\partial z}-v_x\frac{\partial A_s}{\partial x} - v_y\frac{\partial A_s}{\partial y} + v_y\frac{\partial A_y}{\partial z})   
\end{eqnarray}
The first term on the right-hand side of the equation corresponds to the conventional space-charge contribution, which scales as \(1/\gamma^2\). The remaining terms represent additional force components arising from the magnetic fields induced by the transverse and longitudinal currents. 
We denote these two contributions separately as \(F^{\mathrm{sc}}\) and \(F^{\mathrm{r}}\), where \(F^{\mathrm{sc}}\) is the conventional space-charge force, written as:
\begin{eqnarray}
F_x^{sc} & = & -\frac{q}{\gamma^2}\frac{\partial \phi}{\partial x}  \\
F_y^{sc} & = & -\frac{q}{\gamma^2}\frac{\partial \phi}{\partial y}\\  
F_s^{sc} & = & -\frac{q}{\gamma^2}\frac{\partial \phi}{\partial z}  
\end{eqnarray}
and \(F^{r}\) denotes the remaining magnetic-field contributions to the force, written as:
\begin{eqnarray}
F_x^r & = & q(
    v_y\frac{\partial A_y}{\partial x}-v_y\frac{\partial A_x}{\partial y})  \\
F_y^r & = & q(- v_x\frac{\partial A_y}{\partial x} + v_x\frac{\partial A_x}{\partial y}) \\  
F_s^r & = & q(
    v_x\frac{\partial A_x}{\partial z}-v_x\frac{\partial A_s}{\partial x} - v_y\frac{\partial A_s}{\partial y} + v_y\frac{\partial A_y}{\partial z})   
\end{eqnarray}

For a charged-particle beam under the normal condition \(v_0 \gg v_{x,y}\), \(J_s \gg J_{x,y}\), and \(A_s \gg A_{x,y}\), the space-charge term dominates.  
However, for a highly converging or diverging beam induced by strong focusing or defocusing, the remaining terms may become significant.  
In the next section, we compare the magnitude of these two contributions using a three-dimensional Gaussian density distribution.

\section{Generalized space-charge solver for a 3D converging/diverging Gaussian beam}
In this section, we consider a transversely focused charged-particle beam with a six-dimensional Gaussian distribution in the moving-window coordinates as:
\begin{eqnarray}
    f(x,v_x,y,v_y,z,v_z) & = & Q\frac{1}{2\pi\sqrt{\Delta_x}}\exp{(-\frac{1}{2\Delta_x}(\Sigma_{22}^x x^2-2\Sigma_{12}^x xv_x+\Sigma_{11}^x v_x^2))}\times
    \frac{1}{2\pi\sqrt{\Delta_y}}\exp{(-\frac{1}{2\Delta_y}(\Sigma_{22}^y y^2-2\Sigma_{12}^y yv_y+\Sigma_{11}^y v_y^2))}
    \nonumber \\
    & &  \times \frac{1}{2\pi\sigma_z \sigma_{v_z}}\exp{(-\frac{1}{2}(\frac{z^2}{\sigma_z^2}+\frac{v_z^2}{\sigma_{v_z}^2}))}
\end{eqnarray}
where $Q$ is the total charge, $\Delta_{x,y}=\Sigma_{11}^{x,y}\Sigma_{22}^{x,y}-(\Sigma_{12}^2)^{x,y}$, and $\Sigma_{11}^{x,y}$, 
$\Sigma_{12}^{x,y}$, $\Sigma_{22}^{x,y}$ are elements of the covariance matrix of
the Gaussian distribution in the $x$ or $y$ plane.
Given the above phase space distribution, the charge density distribution $\rho(x,y,z)$ and the
transverse current density distribution $J_{x,y}(x,y,z)$ are:
\begin{eqnarray}
    \rho(x,y,z) & = & \frac{Q}{\sqrt{(2\pi)^3 \Sigma_{11}^x\Sigma_{11}^y\sigma_z^2}}\exp{(-\frac{1}{2}(\frac{x^2}{\Sigma_{11}^x}+\frac{y^2}{\Sigma_{11}^y}+\frac{z^2}{\sigma_z^2}))} \\
J_x(x,y,z) & = & \frac{Q}{\sqrt{(2\pi)^3 \Sigma_{11}^x\Sigma_{11}^y\sigma_z^2}}\frac{\Sigma_{12}^x}{\Sigma_{11}^x}x\exp{(-\frac{1}{2}(\frac{x^2}{\Sigma_{11}^x}+\frac{y^2}{\Sigma_{11}^y}+\frac{z^2}{\sigma_z^2}))} \\
J_y(x,y,z) & = & \frac{Q}{\sqrt{(2\pi)^3 \Sigma_{11}^x\Sigma_{11}^y\sigma_z^2}}\frac{\Sigma_{12}^y}{\Sigma_{11}^y}y\exp{(-\frac{1}{2}(\frac{x^2}{\Sigma_{11}^x}+\frac{y^2}{\Sigma_{11}^y}+\frac{z^2}{\sigma_z^2}))} 
\end{eqnarray}
For the charge and current densities above, the electric potential and the vector potential can be obtained by solving the Poisson-like equations~\ref{ep1}--\ref{ep4} using the Fourier transform method as~\cite{qiang2025}:
\begin{eqnarray}
\phi(x,y,z) & = & \frac{Q}{4\pi\epsilon_0 \sqrt{2\pi}}\int_0^{\infty} du
    \frac{\exp[-\frac{x^2}{2(\Sigma_{11}^x+u)}-\frac{y^2}{2(\Sigma_{11}^y+u)}-\frac{z^2}{2(\sigma_z^2+u/\gamma^2)}]}{\sqrt{(\Sigma_{11}^x+u)(\Sigma_{11}^y+u)(\sigma_z^2+u/\gamma^2) }    }    \\
A_x(x,y,z) & = & \frac{Q}{4\pi\epsilon_0 \sqrt{2\pi}}\frac{\Sigma_{12}^x}{c^2}\int_0^{\infty} du
    \frac{x\exp[-\frac{x^2}{2(\Sigma_{11}^x+u)}-\frac{y^2}{2(\Sigma_{11}^y+u)}-\frac{z^2}{2(\sigma_z^2+u/\gamma^2)}]}{(\Sigma_{11}^x+u)\sqrt{(\Sigma_{11}^x+u)(\Sigma_{11}^y+u)(\sigma_z^2+u/\gamma^2) }    }    \\
A_y(x,y,z) & = & \frac{Q}{4\pi\epsilon_0 \sqrt{2\pi}}\frac{\Sigma_{12}^y}{c^2}\int_0^{\infty} du
    \frac{y\exp[-\frac{x^2}{2(\Sigma_{11}^x+u)}-\frac{y^2}{2(\Sigma_{11}^y+u)}-\frac{z^2}{2(\sigma_z^2+u/\gamma^2)}]}{(\Sigma_{11}^y+u)\sqrt{(\Sigma_{11}^x+u)(\Sigma_{11}^y+u)(\sigma_z^2+u/\gamma^2) }    }    \\
A_s(x,y,z) & = & \frac{Q}{4\pi\epsilon_0 \sqrt{2\pi}}\frac{\beta}{c}\int_0^{\infty} du
    \frac{\exp[-\frac{x^2}{2(\Sigma_{11}^x+u)}-\frac{y^2}{2(\Sigma_{11}^y+u)}-\frac{z^2}{2(\sigma_z^2+u/\gamma^2)}]}{\sqrt{(\Sigma_{11}^x+u)(\Sigma_{11}^y+u)(\sigma_z^2+u/\gamma^2) }    }    
\end{eqnarray}
Using the above potentials, the conventional space-charge forces can be obtained as:
\begin{eqnarray}
    F_{x}^{sc} & = &  \frac{qQ}{\gamma^24\pi\epsilon_0 \sqrt{2\pi}}\int_0^{\infty} du
    \frac{x\exp[-\frac{x^2}{2(\Sigma_{11}^x+u)}-\frac{y^2}{2(\Sigma_{11}^y+u)}-\frac{z^2}{2(\sigma_z^2+u/\gamma^2)}]}{(\Sigma_{11}^x+u)\sqrt{(\Sigma_{11}^x+u)(\Sigma_{11}^y+u)(\sigma_z^2+u/\gamma^2) }    }   \\
       F_{y}^{sc}& = &  \frac{qQ}{\gamma^24\pi\epsilon_0 \sqrt{2\pi}}\int_0^{\infty} du
    \frac{y\exp[-\frac{x^2}{2(\Sigma_{11}^x+u)}-\frac{y^2}{2(\Sigma_{11}^y+u)}-\frac{z^2}{2(\sigma_z^2+u/\gamma^2)}]}{(\Sigma_{11}^y+u)\sqrt{(\Sigma_{11}^x+u)(\Sigma_{11}^y+u)(\sigma_z^2+u/\gamma^2) }    }    \\
       F_{s}^{sc} & = &  \frac{qQ}{\gamma^2 4\pi\epsilon_0 \sqrt{2\pi}}\int_0^{\infty} du
    \frac{z\exp[-\frac{x^2}{2(\Sigma_{11}^x+u)}-\frac{y^2}{2(\Sigma_{11}^y+u)}-\frac{z^2}{2(\sigma_z^2+u/\gamma^2)}]}{(\sigma_z^2+u/\gamma^2)\sqrt{(\Sigma_{11}^x+u)(\Sigma_{11}^y+u)(\sigma_z^2+u/\gamma^2) }    }   
\end{eqnarray}
and the remaining (additional) contributions can be written as:
\begin{eqnarray}
F_x^r & = & \frac{qQ}{4\pi\epsilon_0 \sqrt{2\pi}}\frac{\beta_y}{c}(\Sigma_{12}^x-\Sigma_{12}^y))\int_0^{\infty} du
    \frac{xy\exp[-\frac{x^2}{2(\Sigma_{11}^x+u)}-\frac{y^2}{2(\Sigma_{11}^y+u)}-\frac{z^2}{2(\sigma_z^2+u/\gamma^2)}]}{(\Sigma_{11}^x+u)(\Sigma_{11}^y+u)\sqrt{(\Sigma_{11}^x+u)(\Sigma_{11}^y+u)(\sigma_z^2+u/\gamma^2) }    }    \\   
F_y^r & = & \frac{-qQ}{4\pi\epsilon_0 \sqrt{2\pi}}\frac{\beta_x}{c}(\Sigma_{12}^x-\Sigma_{12}^y))\int_0^{\infty} du
    \frac{xy\exp[-\frac{x^2}{2(\Sigma_{11}^x+u)}-\frac{y^2}{2(\Sigma_{11}^y+u)}-\frac{z^2}{2(\sigma_z^2+u/\gamma^2)}]}{(\Sigma_{11}^x+u)(\Sigma_{11}^y+u)\sqrt{(\Sigma_{11}^x+u)(\Sigma_{11}^y+u)(\sigma_z^2+u/\gamma^2) }    }    \\  
F_s^r & = & \frac{qQ}{4\pi\epsilon_0 \sqrt{2\pi}}( 
\beta_0 \beta_x \int_0^{\infty} du
    \frac{x\exp[-\frac{x^2}{2(\Sigma_{11}^x+u)}-\frac{y^2}{2(\Sigma_{11}^y+u)}-\frac{z^2}{2(\sigma_z^2+u/\gamma^2)}]}{(\Sigma_{11}^x+u)\sqrt{(\Sigma_{11}^x+u)(\Sigma_{11}^y+u)(\sigma_z^2+u/\gamma^2) }    }   \nonumber \\
& & +\beta_0 \beta_y \int_0^{\infty} du
    \frac{y\exp[-\frac{x^2}{2(\Sigma_{11}^x+u)}-\frac{y^2}{2(\Sigma_{11}^y+u)}-\frac{z^2}{2(\sigma_z^2+u/\gamma^2)}]}{(\Sigma_{11}^y+u)\sqrt{(\Sigma_{11}^x+u)(\Sigma_{11}^y+u)(\sigma_z^2+u/\gamma^2) }    }  \nonumber \\
& & -\frac{\beta_x}{c}\Sigma_{12}^x
\int_0^{\infty} du
    \frac{xz\exp[-\frac{x^2}{2(\Sigma_{11}^x+u)}-\frac{y^2}{2(\Sigma_{11}^y+u)}-\frac{z^2}{2(\sigma_z^2+u/\gamma^2)}]}{(\Sigma_{11}^x+u)(\sigma_z^2+u/\gamma^2)\sqrt{(\Sigma_{11}^x+u)(\Sigma_{11}^y+u)(\sigma_z^2+u/\gamma^2) }    }  \nonumber \\
& & -
\frac{\beta_y}{c}\Sigma_{12}^y
\int_0^{\infty} du
    \frac{yz\exp[-\frac{x^2}{2(\Sigma_{11}^x+u)}-\frac{y^2}{2(\Sigma_{11}^y+u)}-\frac{z^2}{2(\sigma_z^2+u/\gamma^2)}]}{(\Sigma_{11}^y+u)(\sigma_z^2+u/\gamma^2)\sqrt{(\Sigma_{11}^x+u)(\Sigma_{11}^y+u)(\sigma_z^2+u/\gamma^2) }    }  )
\end{eqnarray}
where \(\beta_0=v_0/c\), and \(\beta_{x,y}=v_{x,y}/c\).
From the above equations, we see that the conventional space-charge term scales down as \(1/\gamma^2\), while the remaining force contributions are not explicitly dependent on \(\gamma\). The impact of a focusing (converging) beam is governed by the value of the correlation element \(\Sigma_{12}^{x,y}\).

The covariance matrix elements \(\Sigma_{11}\), \(\Sigma_{12}\), and \(\Sigma_{22}\) can be obtained from an initial upright six-dimensional Gaussian distribution subjected to transverse focusing and a drift for a time \(t\), as:
\begin{eqnarray}
    \Sigma_{11}^x(t) & = & \Sigma_{11}^{x0} + 2 \Sigma_{12}^{x0} t + \Sigma_{22}^{x0}t^2 \\
    \Sigma_{12}^x(t) & = & \Sigma_{12}^{x0}+\Sigma_{22}^{x0} t \\
    \Sigma_{22}^x(t) & = & \Sigma_{22}^{x0}  
\end{eqnarray}
where \(\Sigma^{x0}\) are the covariance-matrix elements after the transverse focusing.  
For two-dimensional transverse focusing, the same equations are applied to the vertical \(y\) dimension.

Assume an initial upright six-dimensional Gaussian distribution with standard deviations
\(\sigma_x\), \(\sigma_{v_x}\), \(\sigma_y\), \(\sigma_{v_y}\), \(\sigma_z\), and \(\sigma_{v_z}\).
The beam then undergoes transverse focusing with a transfer matrix in the horizontal \(x\) dimension given by:
\begin{equation}
	M_x  =   \left( \begin{array}{cc}
			\cos(\sqrt{K_x}L) & \frac{\sin(\sqrt{K_x}L)}{v_0\sqrt{K_x}} \\
			-v_0\sqrt{K_x}\sin(\sqrt{K_x}L) & \cos(\sqrt{K_x}L) 
	\end{array} \right)
\end{equation}
where $K_x$ is the normalized magnet focusing strength. For a quadrupole, $K_x=qG/mc\gamma \beta$ and $p_0 = \gamma \beta_0$.
A similar transfer matrix can be assumed for the vertical y dimension
for transverse constant focusing.
The covariance matrix elements after the focusing can be written as:
\begin{eqnarray}
    \Sigma_{11}^{x0} & = & (\sigma_x \cos{(\sqrt{K_x}L)})^2+
    (\frac{\sigma_{v_x}\sin{(\sqrt{K_x}L})}{v_0\sqrt{K_x} })^2  \\
  \Sigma_{12}^{x0} & = & \frac{\sigma_{v_x}^2\sin{(\sqrt{K_x}L})\cos{(\sqrt{K_x}L)}}{v_0\sqrt{K_x}}-\sigma_x^2 v_0\sqrt{K_x} \sin{(\sqrt{K_x}L})\cos{(\sqrt{K_x}L)}
   \\   
\Sigma_{22}^{x0} & = & (\sigma_{v_x} \cos{(\sqrt{K_x}L)})^2+
    (\sigma_{x}v_0\sqrt{K_x}\sin{(\sqrt{K_x}L}))^2 
\end{eqnarray}
Given the condition \(v_0\sqrt{K}\,\sigma_x \gg \sigma_{v_x}\), the above expressions can be approximated as:
\begin{eqnarray}
    \Sigma_{11}^{x0} & \approx& (\sigma_x \cos{(\sqrt{K_x}L)})^2 \\
  \Sigma_{12}^{x0} & \approx & -\sigma_x^2 v_0\sqrt{K_x} \sin{(\sqrt{K_x}L})\cos{(\sqrt{K_x}L)}
   \\   
\Sigma_{22}^{x0} & \approx & 
    (\sigma_{x}v_0\sqrt{K_x}\sin{(\sqrt{K_x}L}))^2 
\end{eqnarray}

For the convenience of comparison, we introduce a variable $w^2 = \frac{1}{1 + t/\Sigma_{11}^x}$ in the above integrals. The conventional space-charge forces can then be rewritten as:
\begin{eqnarray}
\label{fconv1}
    F_{x}^{sc}(x,y,z) & = & \frac{qQ2Ax}{\gamma^24\pi\epsilon_0\sqrt{2\pi}\sigma_z\Sigma_{11}^x}\int_0^{1} dw
    \frac{w^2\exp[-\frac{1}{2}w^2x'^2-\frac{1}{2}w^2y'^2/(A^2+(1-A^2)w^2)-\frac{1}{2}w^2z'^2/(B^2+(1-B^2)w^2)]}{\sqrt{(A^2+(1-A^2)w^2)(B^2+(1-B^2)w^2) }    }  \\
        F_{y}^{sc}(x,y,z) & = & \frac{qQ2Ay}{\gamma^2 4\pi\epsilon_0\sqrt{2\pi}\sigma_z\Sigma_{11}^y}\int_0^{1} dw
    \frac{w^2\exp[-\frac{1}{2}w^2x'^2-\frac{1}{2}w^2y'^2/(A^2+(1-A^2)w^2)-\frac{1}{2}w^2z'^2/(B^2+(1-B^2)w^2)]}{(A^2+(1-A^2)w^2)\sqrt{(A^2+(1-A^2)w^2)(B^2+(1-B^2)w^2) }    }  \\
      F_{z}^{sc} (x,y,z) & = & \frac{qQ2Az}{\gamma^2 4\pi\epsilon_0\sqrt{2\pi}\sigma_z\sigma_z^2}\int_0^{1} dw
    \frac{w^2\exp[-\frac{1}{2}w^2x'^2-\frac{1}{2}w^2y'^2/(A^2+(1-A^2)w^2)-\frac{1}{2}w^2z'^2/(B^2+(1-B^2)w^2)]}{(B^2+(1-B^2)w^2)\sqrt{(A^2+(1-A^2)w^2)(B^2+(1-B^2)w^2) }    }  
    \label{fconv3}
\end{eqnarray}
where the aspect ratio $A = \sqrt{\Sigma_{11}^x / \Sigma_{11}^y}$, $B=\sqrt{\Sigma_{11}^x}/(\gamma \sigma_z)$, $x' = x / \sqrt{\Sigma_{11}^x}$, $y' = y / \sqrt{\Sigma_{11}^y}$, and $z'=z/\sigma_z$. This transformation converts the integral from an infinite domain to a finite range between $0$ and $1$. 
The remaining forces can be rewritten as:
\begin{eqnarray}
\label{frest1}
    F_{x}^{r}(x,y,z) & = & \frac{qQ2Axy}{4\pi\epsilon_0\sqrt{2\pi}\sigma_z\Sigma_{11}^x\Sigma_{11}^y} \frac{\beta_y}{c}(\Sigma_{12}^x-\Sigma_{12}^y)) \times \nonumber \\
    & & \int_0^{1} dw
    \frac{w^4\exp[-\frac{1}{2}w^2x'^2-\frac{1}{2}w^2y'^2/(A^2+(1-A^2)w^2)-\frac{1}{2}w^2z'^2/(B^2+(1-B^2)w^2)]}{(A^2+(1-A^2)w^2)\sqrt{(A^2+(1-A^2)w^2)(B^2+(1-B^2)w^2) }    }  \\
        F_{y}^{r}(x,y,z) & = & \frac{-qQ2Axy}{ 4\pi\epsilon_0\sqrt{2\pi}\sigma_z\Sigma_{11}^x\Sigma_{11}^y}\frac{\beta_x}{c}(\Sigma_{12}^x-\Sigma_{12}^y)) \times \nonumber \\
        & & \int_0^{1} dw
    \frac{w^4\exp[-\frac{1}{2}w^2x'^2-\frac{1}{2}w^2y'^2/(A^2+(1-A^2)w^2)-\frac{1}{2}w^2z'^2/(B^2+(1-B^2)w^2)]}{(A^2+(1-A^2)w^2)\sqrt{(A^2+(1-A^2)w^2)(B^2+(1-B^2)w^2) }    }  \\
      F_{z}^{r} (x,y,z) & = & 
      \frac{qQ2A}{4\pi\epsilon_0\sqrt{2\pi}\sigma_z}(
    \frac{x\beta_0\beta_x}{\Sigma_{11}^x}\int_0^{1} dw
    \frac{w^2\exp[-\frac{1}{2}w^2x'^2-\frac{1}{2}w^2y'^2/(A^2+(1-A^2)w^2)-\frac{1}{2}w^2z'^2/(B^2+(1-B^2)w^2)]}{\sqrt{(A^2+(1-A^2)w^2)(B^2+(1-B^2)w^2) }    }  \nonumber \\
    & & + \frac{y\beta_0\beta_y}{\Sigma_{11}^y}\int_0^{1} dw
    \frac{w^2\exp[-\frac{1}{2}w^2x'^2-\frac{1}{2}w^2y'^2/(A^2+(1-A^2)w^2)-\frac{1}{2}w^2z'^2/(B^2+(1-B^2)w^2)]}{(A^2+(1-A^2)w^2)\sqrt{(A^2+(1-A^2)w^2)(B^2+(1-B^2)w^2) }    }  \nonumber \\
   & &  - \frac{\beta_x}{c}\frac{xz\Sigma_{12}^x}{\sigma_z^2\Sigma_{11}^x}\int_0^{1} dw
    \frac{w^4\exp[-\frac{1}{2}w^2x'^2-\frac{1}{2}w^2y'^2/(A^2+(1-A^2)w^2)-\frac{1}{2}w^2z'^2/(B^2+(1-B^2)w^2)]}{(B^2+(1-B^2)w^2)\sqrt{(A^2+(1-A^2)w^2)(B^2+(1-B^2)w^2) }    }  \nonumber \\
    & & - \frac{\beta_y}{c}\frac{yz\Sigma_{12}^y}{\sigma_z^2\Sigma_{11}^y}\int_0^{1} dw
    \frac{w^4\exp[-\frac{1}{2}w^2x'^2-\frac{1}{2}w^2y'^2/(A^2+(1-A^2)w^2)-\frac{1}{2}w^2z'^2/(B^2+(1-B^2)w^2)]}{(A^2+(1-A^2)w^2)(B^2+(1-B^2)w^2)\sqrt{(A^2+(1-A^2)w^2)(B^2+(1-B^2)w^2) } }  )
    \label{frest3}
\end{eqnarray}

The integrals in Eqs.~\ref{fconv1}--\ref{frest3} are dimensionless and can be directly compared.

For the transverse forces, the integral in Eq.~\ref{frest1} (the remaining-force contribution) is smaller than the corresponding one in Eq.~\ref{fconv1}, since \(w<1\). Therefore, if the transverse remaining forces are to be comparable to the conventional transverse space-charge forces, this suggests that:
\begin{equation}
    \frac{y}{\Sigma_{11}^y}\,\frac{\beta_y}{c}\,\Sigma_{12}^x \sim \frac{1}{\gamma^2}
\end{equation}
Assuming that the transverse correlation arises from the focusing discussed above, this leads to:
\begin{eqnarray}
\sigma_y \sigma_{\beta_y} \sqrt{K_y} \sim  \frac{1}{\gamma^2} 
\end{eqnarray}
or
\begin{eqnarray}
    \epsilon_{ny} \sqrt{K_y} \sim \frac{1}{\gamma}
\end{eqnarray}
where \(\epsilon_{ny}\) is the normalized vertical emittance.  

Given that \(\sqrt{K_y}\sim O(1)\) and that the transverse normalized emittance is \(\epsilon_{ny}\sim 1\times 10^{-6}\), this estimate implies that \(\gamma\) needs to be on the order of \(10^6\) for the remaining transverse force to become comparable to the conventional transverse space-charge force.

In the longitudinal direction, the conventional space-charge force scales down as \(1/\gamma^2\). The remaining longitudinal force comprises four terms. The first two terms are dominant, since the other two are proportional to \(\beta/c\).  

For the remaining force to be comparable to the conventional space-charge force, this suggests that
\begin{eqnarray}
     \sigma_{\beta_x}  \sim \frac{I}{\gamma^2}
\end{eqnarray}
or 
\begin{eqnarray}
    \sigma_{\gamma \beta_x}   \sim  \frac{I}{\gamma}
\end{eqnarray}
where I is the ratio of the dimensionless integral in $F_z^{sc}$ and $F_z^r$ and is roughly a factor of 10.
Taking \(\sigma_{\gamma \beta_x}\sim 10^{-3}\), this estimate suggests that \(\gamma \sim 10^4\) for the remaining force to be comparable to the conventional space-charge term.

\begin{figure}[!htb]
   \centering
   \includegraphics*[angle=0,width=225pt]{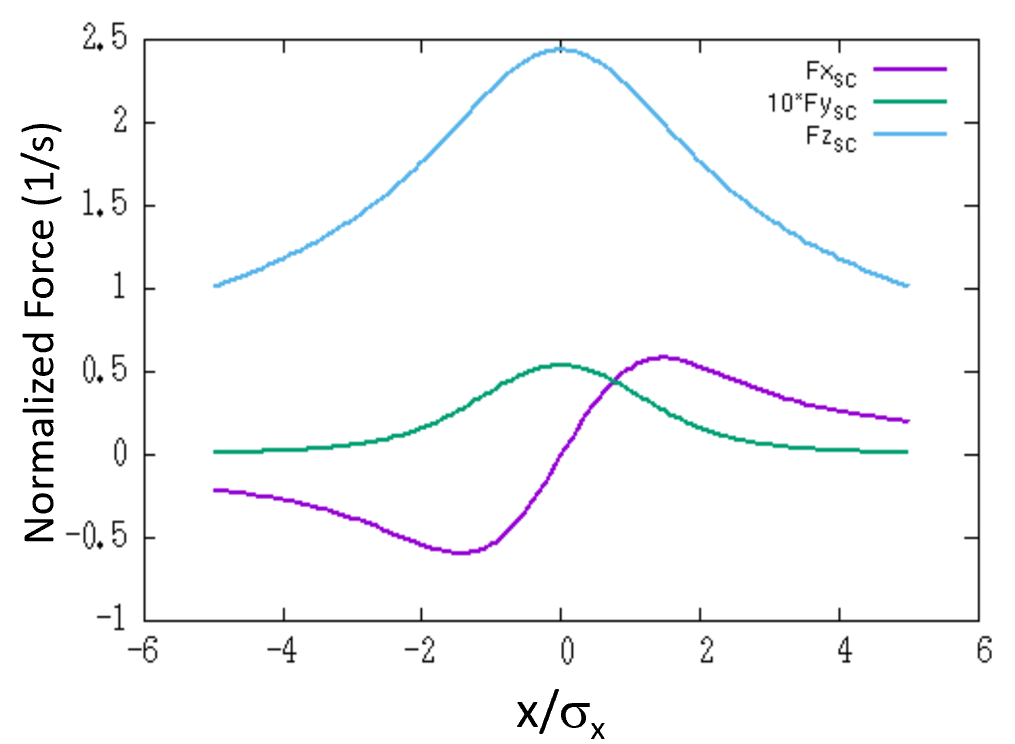}
    \includegraphics*[angle=0,width=230pt]{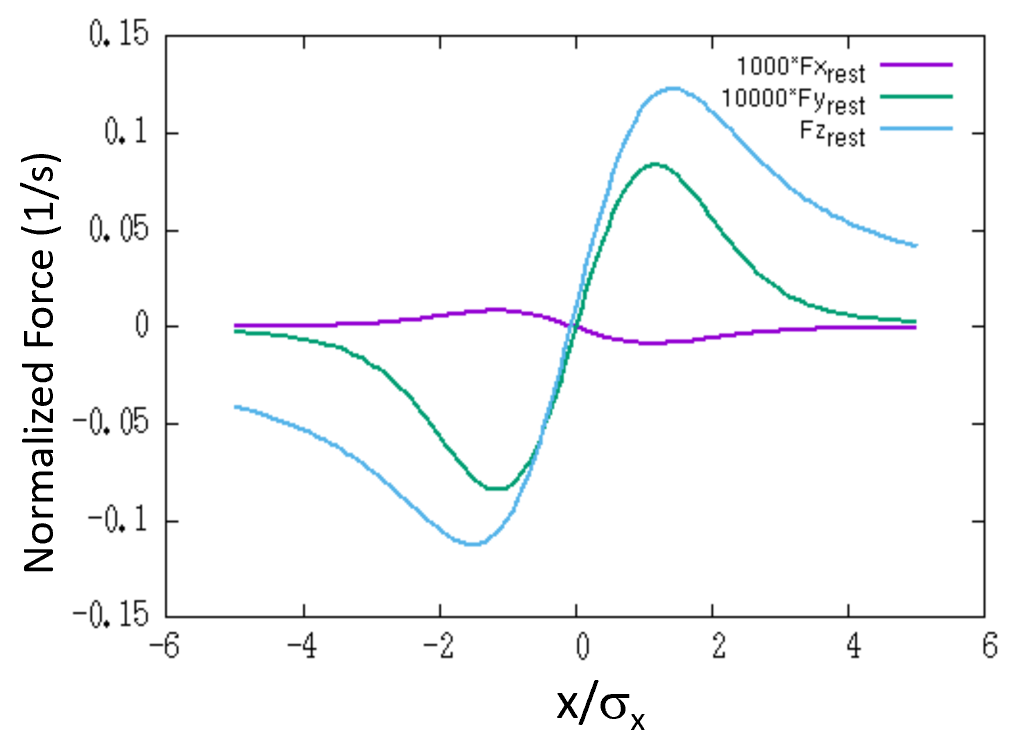}
   \caption{Normalized conventional space-charge forces (left) and
   the rest of forces (right) as a function of horizontal coordinate at $p_x=p_y=0.001$ and $y=0.078\sigma_y$, $z=0.078\sigma_z$ inside a 100 MeV 1 nC electron beam with a 3D converging Gaussian density distribution.}
   \label{Fvsx100mev}
\end{figure}

In the following, we use a concrete example to illustrate the comparison between the conventional space-charge forces and the remaining (additional) space-charge forces.

We consider a three-dimensional electron beam with a Gaussian density distribution characterized by
\(\sigma_x = 1~\mathrm{mm}\), \(\sigma_y = 0.5~\mathrm{mm}\), and \(\sigma_z = 0.1~\mathrm{mm}\).
The transverse covariance elements are taken as
\(\Sigma_{12}^x = -1000~\mathrm{m}\) and \(\Sigma_{12}^y = -500~\mathrm{m}\) (in units of \(\mathrm{m} / \mathrm{s}\)).
Assuming transverse momenta (normalized by \(mc\)) \(p_x=p_y=0.001\), Fig.~\ref{Fvsx100mev} shows the normalized conventional space-charge forces (normalized by \(10^6 mc\)) and the remaining force contributions as a function of the horizontal coordinate at
\(y=0.078\,\sigma_y\) and \(z=0.078\,\sigma_z\).
The beam corresponds to a \(100~\mathrm{MeV}\), \(1~\mathrm{nC}\) electron beam with a three-dimensional converging Gaussian density distribution.

It is seen that the conventional transverse space-charge forces are about three orders of magnitude larger than the remaining transverse forces.
The conventional longitudinal space-charge force is approximately a factor of ten larger than the remaining longitudinal force.

\begin{figure}[!htb]
   \centering
   \includegraphics*[angle=0,width=230pt]{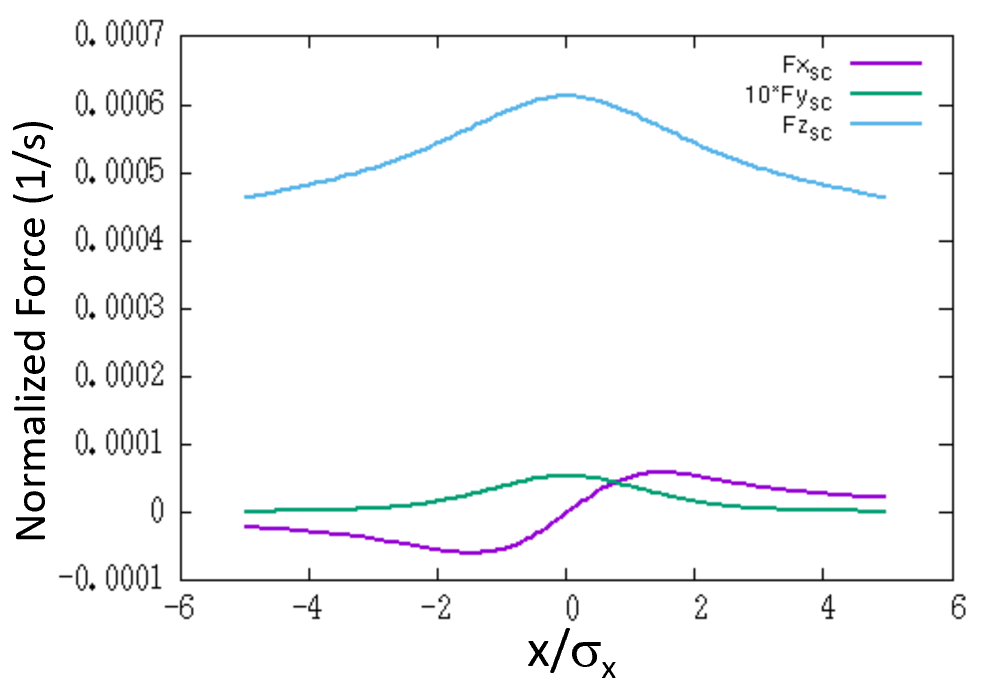}
    \includegraphics*[angle=0,width=230pt]{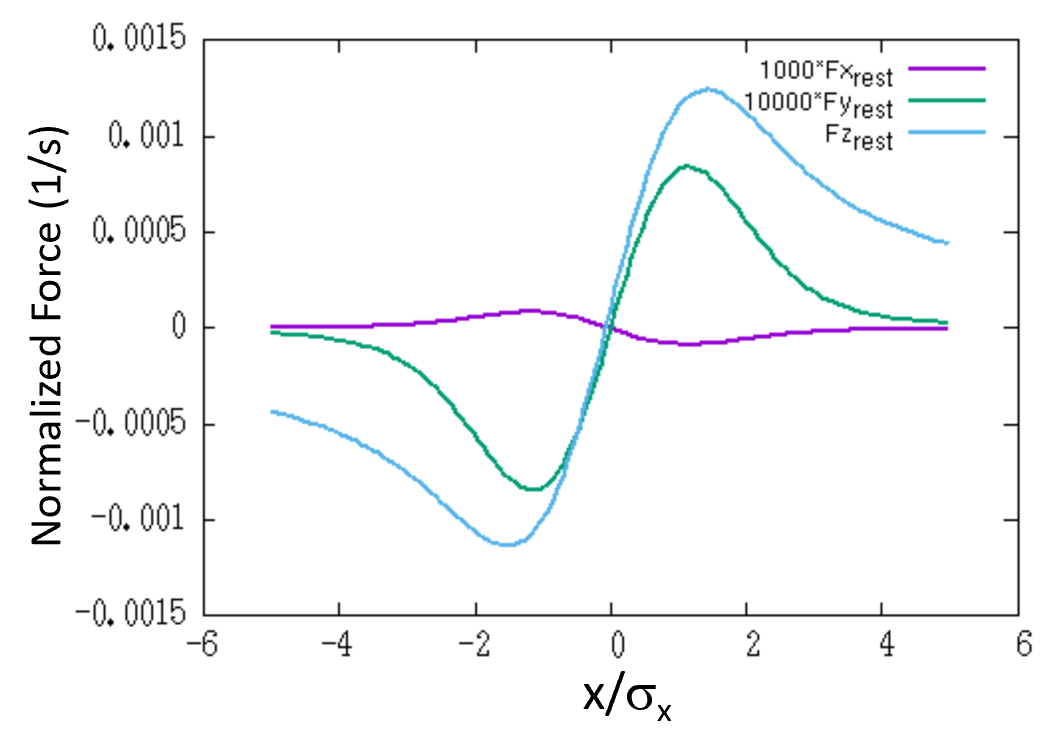}
   \caption{Normalized conventional space-charge forces (left) and
   the rest of forces (right) as a function of horizontal coordinate at $p_x=p_y=0.001$ and $y=0.078\sigma_y$, $z=0.078\sigma_z$ inside a 10 GeV 1 nC electron beam with a 3D converging Gaussian density distribution.}
    \label{Fvsx10gev}
\end{figure}

The conventional space-charge forces scale down as \(1/\gamma^2\). Figure~\ref{Fvsx10gev} shows the normalized conventional space-charge forces and the remaining (additional) force contributions as a function of the horizontal coordinate, evaluated at
\(y=0.078\,\sigma_y\) and \(z=0.078\,\sigma_z\),
inside a \(10~\mathrm{GeV}\), \(1~\mathrm{nC}\) electron beam with a three-dimensional converging Gaussian density distribution.

It is seen that the conventional transverse space-charge forces are still about two orders of magnitude larger than the remaining transverse forces. However, the conventional longitudinal space-charge force becomes smaller than the remaining longitudinal force. In the present example we keep the product \(\gamma\beta\) fixed, implying that \(\beta\) decreases as \(\gamma\) increases.

\section{Generalized space-charge solver for an arbitrary 
density distribution beam}

For an arbitrary density distribution, the Poisson-like equations~\ref{ep1}--\ref{ep4} subject to open boundary conditions can be solved using a Green's function method similar to that used in conventional space-charge solvers~\cite{impact}.

The solution of the above Poisson-like equation under the three-dimensional free-space open boundary condition can be written as:
\begin{eqnarray}
\phi(x,y,z) & = &
\frac{1}{4 \pi \epsilon_0}\int \int \int G(x-x',y-y',z-z') 
\rho(x',y', z') \ dx' dy' dz'  \\
A_x(x,y,z) & = &
\frac{1}{4 \pi c^2 \epsilon_0}\int \int \int G(x-x',y-y',z-z') 
J_x(x',y', z') \ dx' dy' dz'  \\
A_y(x,y,z) & = &
\frac{1}{4 \pi c^2 \epsilon_0}\int \int \int G(x-x',y-y',z-z') 
J_y(x',y', z') \ dx' dy' dz'  \\
A_z(x,y,z) & = &
\frac{1}{4 \pi c^2 \epsilon_0}\int \int \int G(x-x',y-y',z-z') 
J_z(x',y', z') \ dx' dy' dz'  
\end{eqnarray}
where the Green's function $G$ 
is given by:
\begin{equation}
   G(x-x',y-y',z-z') = \frac{\gamma}{\sqrt{(x-x')^2+(y-y')^2+\gamma^2(z-z')^2}}
   \label{green}
\end{equation}

To compute the above convolution integrals numerically, we define a computational domain covering the beam with ranges \((0,L_x)\), \((0,L_y)\), and \((0,L_z)\), and discretize each dimension using \(N_x\), \(N_y\), and \(N_z\) grid points, respectively.

Instead of using the Green's function in Eq.~\ref{green} directly, we employ an integrated Green's function method. This approach can effectively handle the anisotropic computational domain, particularly for large aspect ratios.

Specifically, the integral over the entire computational domain is decomposed into a sum of \(N_x\times N_y\times N_z\) small cell integrals, with each grid point located at the center of a cell. Taking the electric potential as an example, the numerical solution at each grid point can be written as:
\begin{eqnarray}
{\phi}(x_i,y_j,z_k) & = & \frac{1}{4 \pi \epsilon_0}  
\sum_{i'=1}^{N_x} \sum_{j'=1}^{N_y} \sum_{k'=1}^{N_z} 
\int_{x_{i'}-h_x/2}^{x_{i'}+h_x/2} dx'
\int_{y_{j'}-h_y/2}^{y_{j'}+h_y/2} dy' \int_{z_{k'}-h_z/2}^{z_{k'}+h_z/2} dz' \times \nonumber \\
& & G(x_i-x',y_j-y',z_k-z') 
{ \rho}(x',y',z') 
\end{eqnarray}
where $h_x = L_x/(N_x-1)$, $h_y = L_y/(N_y-1)$, and $h_z = L_x/(N_z-1)$.
If we assume that the charge density is constant within
each cell centered at the grid point $(x_i,y_j,z_k)$, i.e.
$\rho(x',y',z')=\rho(x_i,y_j,z_k)$,
from the above equation, the electric potential on this grid point can be
approximated as:
\begin{eqnarray}
{\phi}(x_i,y_j,z_k) & = & \frac{1}{4 \pi \epsilon_0}  
\sum_{i'=1}^{N_x} \sum_{j'=1}^{N_y} \sum_{k'=1}^{N_z} 
{\bar G}(x_i-x_{i'},y_j-y_{j'},z_k-z_{k'}) 
{ \rho}(x_{i'},y_{j'},z_{k'}) \nonumber \\
	& & 
 \label{phiint}
\end{eqnarray}
where $x_i = (i-1)h_x$, $y_j = (j-1)h_y$, and $z_k = (k-1) h_z$,
and the effective Green function ${\bar G}$ is given as: 
\begin{eqnarray}
{\bar G}(x_i-x_{i'},y_j-y_{j'},z_k-z_{k'}) & = & \int_{x_{i'}-h_x/2}^{x_{i'}+h_x/2} dx'
\int_{y_{j'}-h_y/2}^{y_{j'}+h_y/2} dy' \int_{z_{k'}-h_z/2}^{z_{k'}+h_z/2} dz' \times \nonumber \\
& &  G(x_i-x',y_j-y',z_k-z') 
\label{geff}
\end{eqnarray}
where $h_x$, $h_y$, and $h_z$ are cell size in each dimension respectively. 
The above integral can be calculated analytically in a closed form for the Green's function given in Eq.~\ref{green} as ~\cite{qiang2007b}:
\begin{eqnarray}
\bar{G}(x,y,z) & = & f(x+\frac{h_x}{2},y+\frac{h_y}{2},\gamma(z+\frac{h_z}{2})) -
f(x+\frac{h_x}{2},y+\frac{h_y}{2},\gamma(z-\frac{h_z}{2})) + \nonumber \\
& & f(x-\frac{h_x}{2},y-\frac{h_y}{2},\gamma(z+\frac{h_z}{2})) -
f(x-\frac{h_x}{2},y-\frac{h_y}{2},\gamma(z-\frac{h_z}{2})) + \nonumber \\
& & f(x+\frac{h_x}{2},y-\frac{h_y}{2},\gamma(z-\frac{h_z}{2})) -
f(x+\frac{h_x}{2},y-\frac{h_y}{2},\gamma(z+\frac{h_z}{2})) + \nonumber \\
& & f(x-\frac{h_x}{2},y+\frac{h_y}{2},\gamma(z-\frac{h_z}{2})) -
f(x-\frac{h_x}{2},y+\frac{h_y}{2},\gamma(z+\frac{h_z}{2}))
\label{gfunc}
\end{eqnarray}
where 
\begin{eqnarray}
f(x,y,z) =  
yz \ln(x+r) + 
xz \ln(y+r) + 
xy \ln(z+r) -  \nonumber \\
\frac{z^2}{2}\arctan(\frac{xy}{zr}) 
-\frac{y^2}{2}\arctan(\frac{xz}{yr})-
	\frac{x^2}{2}\arctan(\frac{yz}{xr})
 \label{ffunc}
\end{eqnarray}
where $r = \sqrt{x^2+y^2+z^2}$.

The summations appearing in the logarithms of the above expressions may overflow due to cancellation errors in certain extreme cases.
For example, for a very high-energy electron beam (of order \(100~\mathrm{GeV}\)), the longitudinal bunch length in the beam frame can become much larger than the transverse beam size. In this regime, one may have \(|z|\simeq r\), which leads to strong cancellation in the term \(z+r\). When \(z<0\), \(z+r\) can become very small, and \(\ln(z+r)\) may overflow numerically. In such cases, the alternative equivalent expressions can be used to avoid numerical overflow~\cite{qiang2024}.

With the effective Green's function \(\bar{G}\), the summation in Eq.~\ref{phiint} can be computed efficiently using an FFT method~\cite{impact-t,hockney}. The same integrated Green's function and the FFT approach can be applied to compute the vector potentials by replacing the charge density with the corresponding current densities of the charged-particle beam.

As a test of the numerical solver, we consider the same three-dimensional converging Gaussian electron beam at an energy of about \(10~\mathrm{GeV}\), and compute the generalized space-charge forces using both the numerical solver and the semi-analytical Gaussian solver.  

\begin{figure}[!htb]
   \centering
   \includegraphics*[angle=0,width=250pt]{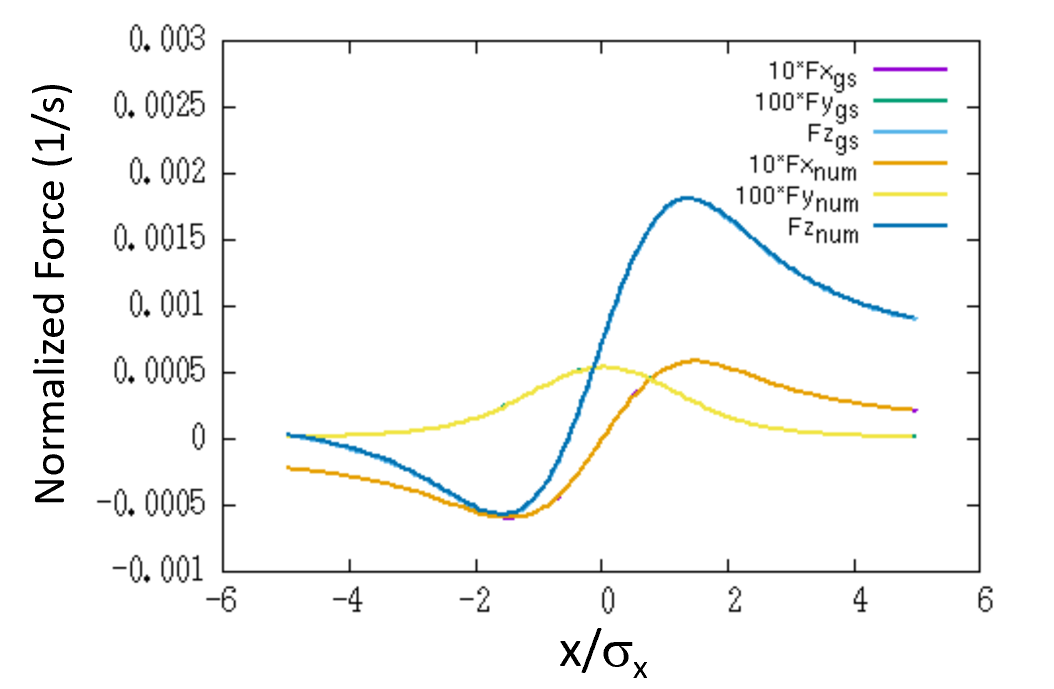}
   \caption{Normalized space-charge forces from Gaussian solver and numerical solver as a function of horizontal coordinate at $p_x=p_y=0.001$ and $y=0.078\sigma_y$, $z=0.078\sigma_z$ inside a 10 GeV 1 nC electron beam with a 3D converging Gaussian density distribution.}
   \label{Fexcmp}
\end{figure}

Figure~\ref{Fexcmp} compares the normalized forces obtained from the two approaches. The two solvers show excellent agreement.

\section{Application examples}

In this section, we apply the generalized three-dimensional space-charge solver to three cases to study space-charge effects in strongly converging beams.

\begin{figure}[!htb]
   \centering
   \includegraphics*[angle=0,width=230pt]{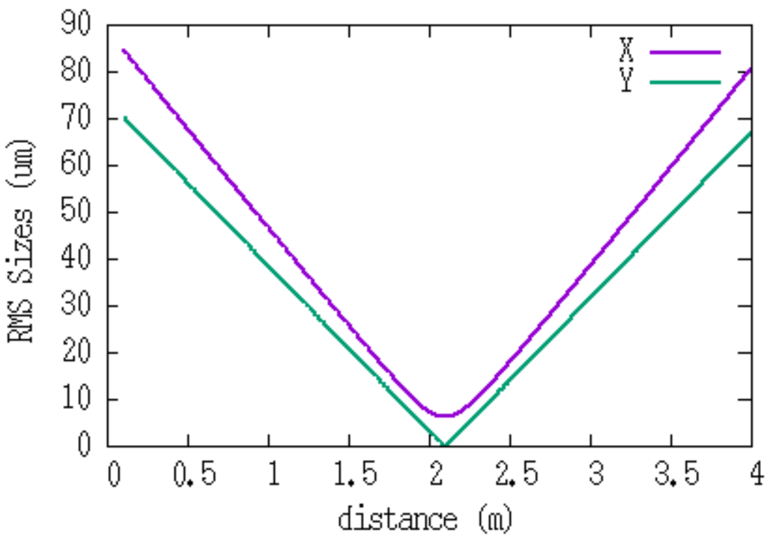}
    \includegraphics*[angle=0,width=225pt]{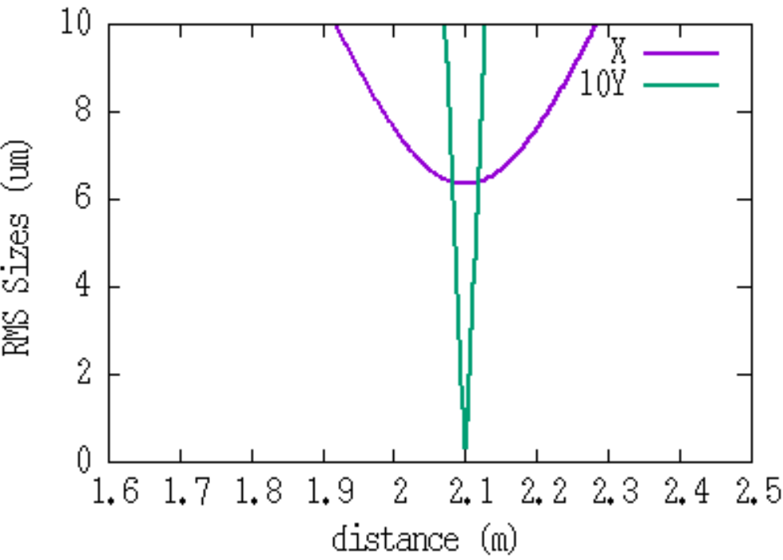}
   \caption{Transverse RMS size evolution inside the FCCee interaction region drift (left) and zoom-in plot (right).}
   \label{trmsfcc}
\end{figure}

The first case is a converging electron beam inside the FCC-ee interaction region. Figure~\ref{trmsfcc} shows the evolution of the transverse RMS beam size through the interaction-region drift. As expected, both the horizontal and vertical beam sizes are strongly focused to a tiny spot at the interaction point in order to maximize the collider luminosity.
In the vertical dimension, the electron beam converges from an initial size of less than \(100~\mu\mathrm{m}\) to a final size below \(30~\mathrm{nm}\).
\begin{figure}[!htb]
   \centering
   \includegraphics*[angle=0,width=230pt]{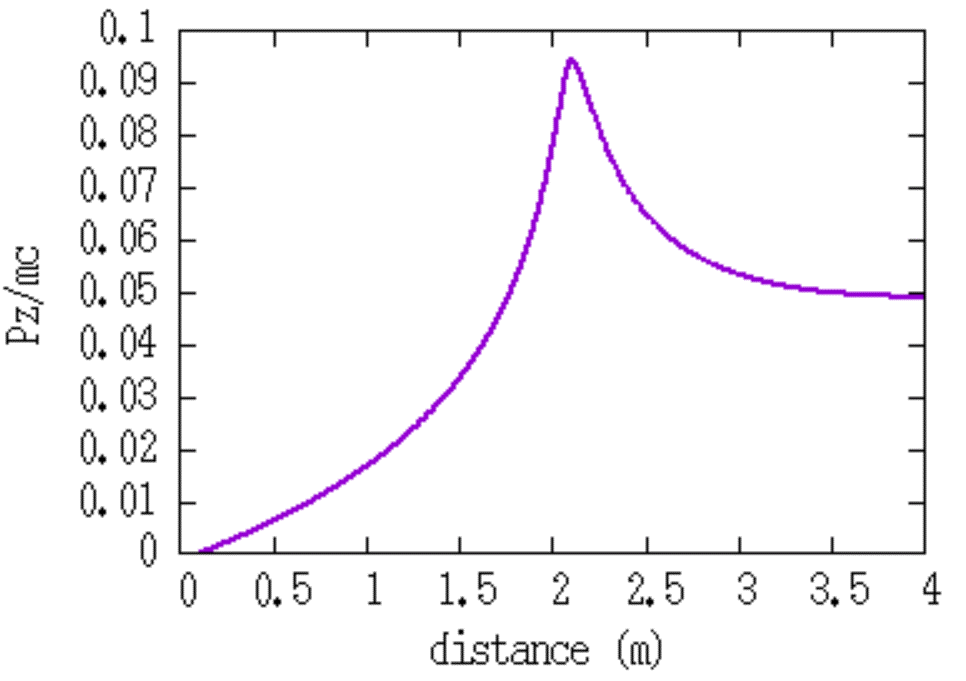}
   \caption{Longitudinal momentum evolution inside the FCCee interaction region.}
   \label{lengfcc}
\end{figure}

Figure~\ref{lengfcc} shows the evolution of the longitudinal RMS momentum inside the interaction region, assuming an initial zero longitudinal momentum spread at the beginning.  
The maximum relative energy spread generated by this momentum spread is about \(10^{-6}\). This is much smaller than the energy spread of \(0.00038\) in the nominal \(45.6~\mathrm{GeV}\) design for the \(Z\)-boson study~\cite{fccee}. This suggests that space-charge effects in the strongly converging beam will not pose a significant problem.

\begin{figure}[!htb]
   \centering
   \includegraphics*[angle=0,width=220pt]{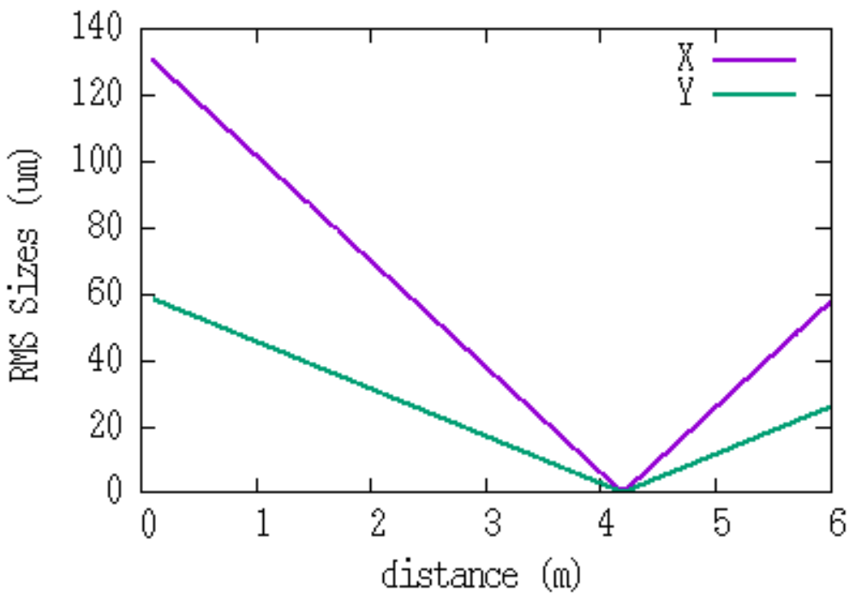}
    \includegraphics*[angle=0,width=240pt]{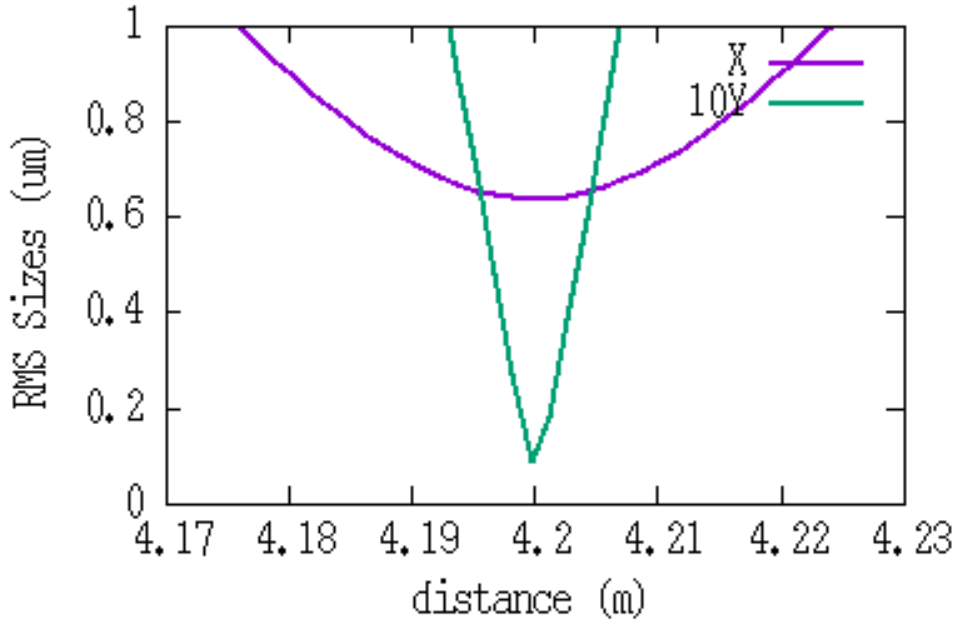}
   \caption{Transverse RMS size evolution inside the ILC interaction region drift (left) and zoom in plot (right).}
   \label{trmsilc}
\end{figure}

The second case is a converging electron beam inside the ILC interaction region. Figure~\ref{trmsilc} shows the evolution of the transverse RMS beam size through the interaction-region drift. As in the FCC-ee case, both the horizontal and vertical beam sizes are strongly focused to tiny spot sizes at the interaction point in order to maximize the collider luminosity. Specifically, the electron beam converges from initial sizes of about \(130~\mu\mathrm{m}\) (horizontal) and \(60~\mu\mathrm{m}\) (vertical) to final sizes of about \(640~\mathrm{nm}\) in the horizontal plane and below \(9~\mathrm{nm}\) in the vertical plane.

\begin{figure}[!htb]
   \centering
   \includegraphics*[angle=0,width=230pt]{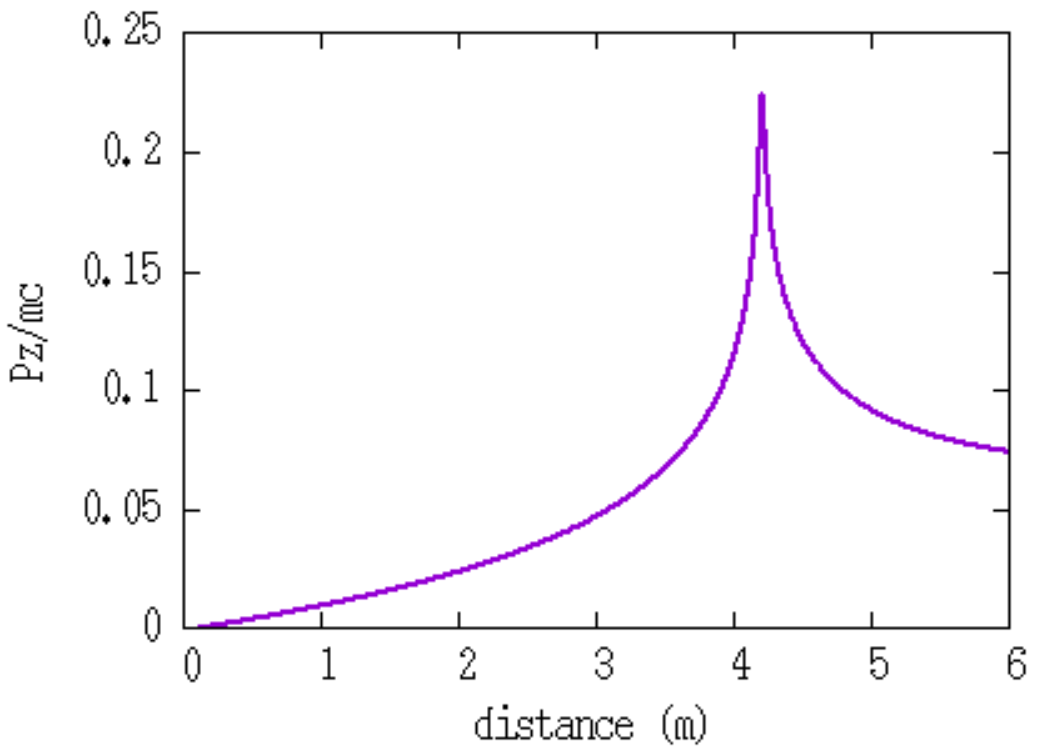}
   \caption{Longitudinal momentum evolution inside the ILC interaction region.}
   \label{lengilc}
\end{figure}

Figure~\ref{lengilc} shows the longitudinal RMS momentum evolution inside the interaction region, assuming an initial zero longitudinal momentum spread at the beginning.  
The maximum relative energy spread generated by this momentum spread is about \(0.5\times 10^{-6}\). This is much smaller than the energy spread of \(0.001\) in the nominal \(250~\mathrm{GeV}\) design~\cite{ilc}.  
This suggests that space-charge effects in the strongly converging beam inside the ILC interaction region will not pose a significant problem either.

In the third case, we study space-charge effects in an electron beam passing through the second bunch compressor in an LCLS-II-HE design~\cite{tor,lcls2he}.

\begin{figure}[!htb]
   \centering
   \includegraphics*[angle=0,width=230pt]{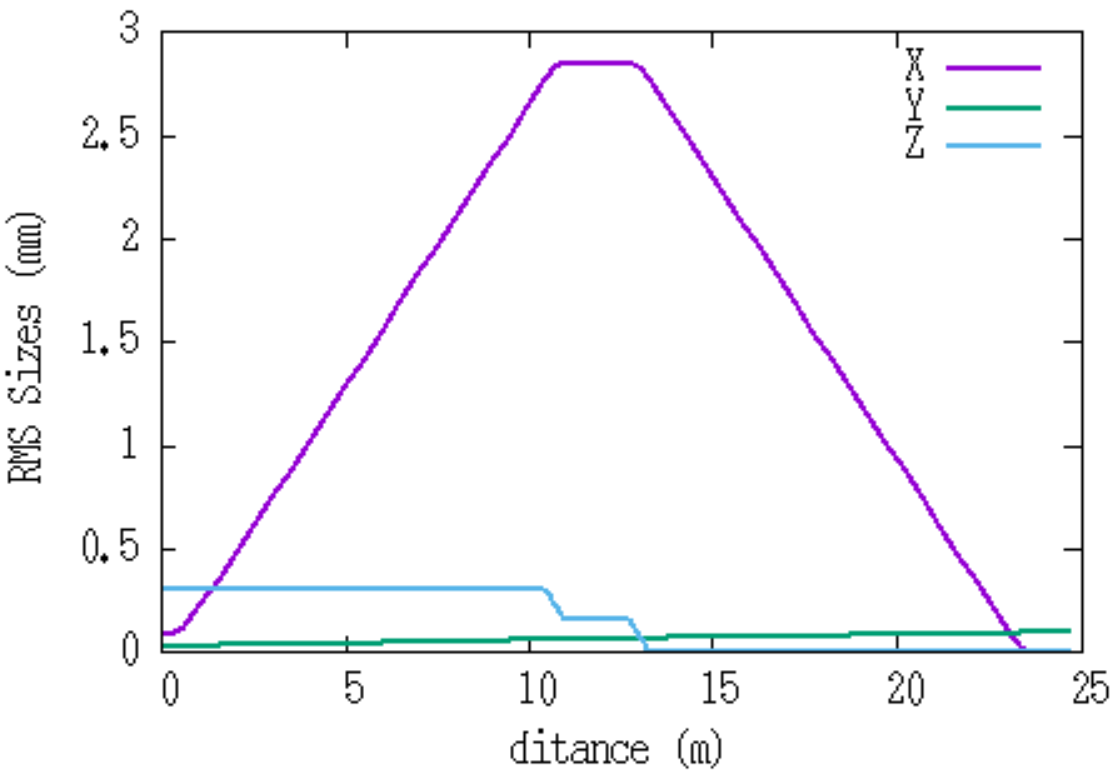}
   \caption{Transverse and longitudinal RMS size evolution through
   the second LCLS-II bunch compressor chicane.}
   \label{rmslcls2}
\end{figure}

Figure~\ref{rmslcls2} shows the evolution of the transverse and longitudinal RMS sizes through the chicane. The electron-beam horizontal size between the 3rd and 4th bending magnets converges from less than \(3~\mathrm{mm}\) to about \(76~\mu\mathrm{m}\). The longitudinal bunch length is compressed by more than a factor of 40 after the 3rd bending magnet, reaching nearly a \(2~\mathrm{kA}\) peak current.

In this study, we use the generalized space-charge model described above to simulate the electron-beam evolution while it converges between the 3rd and 4th bending magnets.

\begin{figure}[!htb]
   \centering
   \includegraphics*[angle=0,width=225pt]{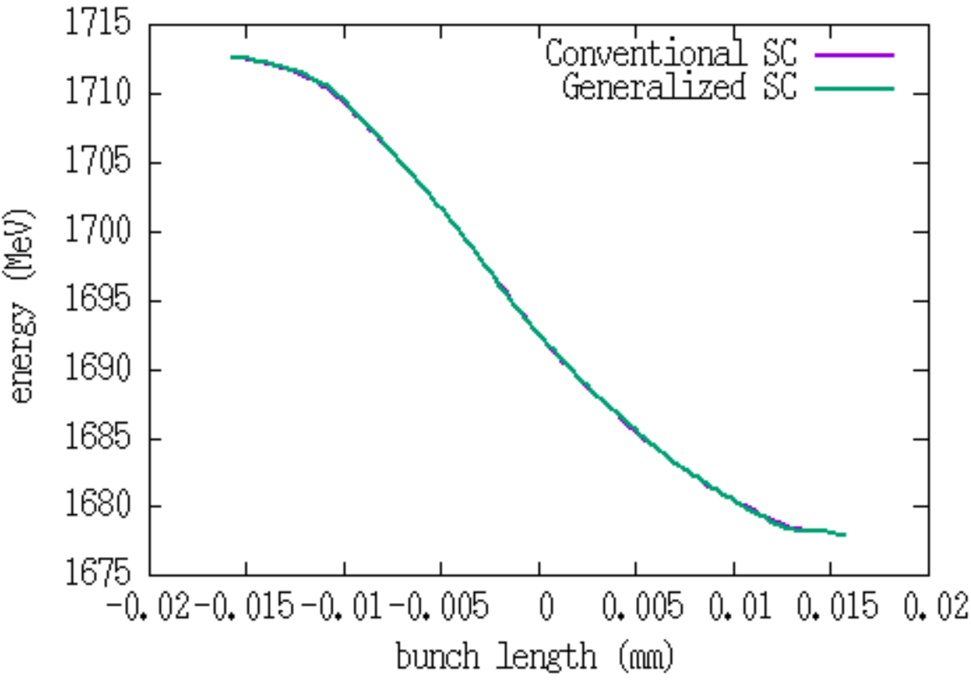}
    \includegraphics*[angle=0,width=230pt]{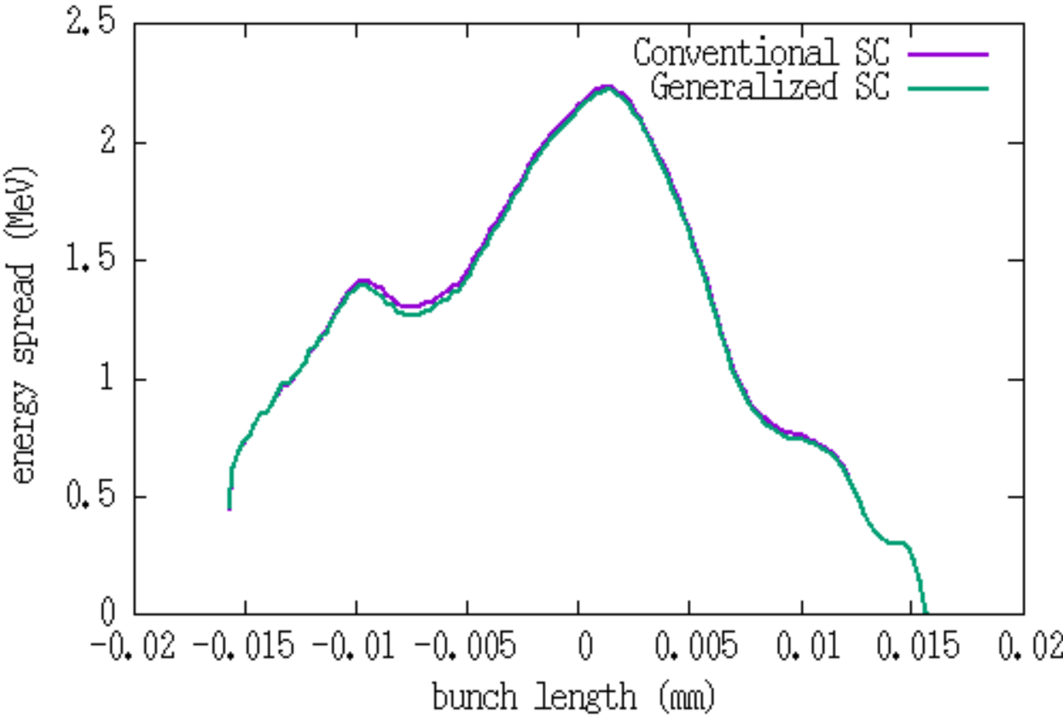}
   \caption{Longitudinal correlated energy spread at the end of the
   drift before the 4th dipole of the chicane (left) and 
   uncorrelated energy spread (right) from the conventional space-charge solver
   and from the generalized space-charge solver.}
   \label{englcls2}
\end{figure}

\begin{figure}[!htb]
   \centering
   \includegraphics*[angle=0,width=230pt]{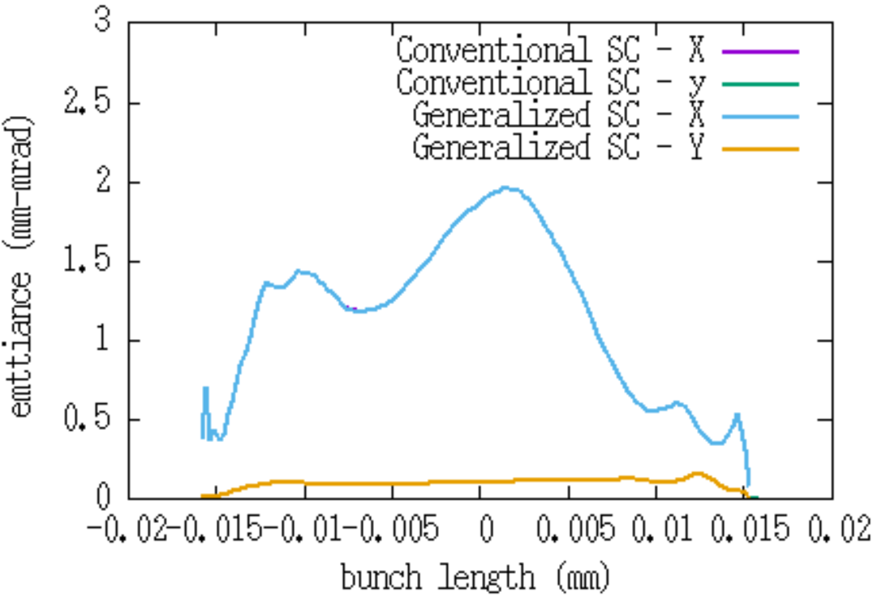}
 \caption{Transverse slice emittance at the end of the
   drift before the 4th dipole of the chicane from the conventional space-charge solver
   and from the generalized space-charge solver.}
   \label{semtlcls2}
\end{figure}

Figure~\ref{englcls2} shows the final correlated and uncorrelated energy spread at the end of the drift for simulations using the conventional space-charge model and the generalized space-charge model. The energy spreads predicted by the two models nearly overlap.

Figure~\ref{semtlcls2} shows the final transverse slice emittance at the end of the drift for simulations using the conventional space-charge model and the generalized space-charge model. 
The slice emittances from the two models lie on top of each other. This suggests that the remaining forces arising from the converging currents do not produce a noticeable impact at this energy of \(1.7~\mathrm{GeV}\).

\section{Conclusions}

In this paper, we present a generalized space-charge solver for three-dimensional converging or diverging beams. Strong focusing or defocusing can generate significant transverse current. The generalized solver includes contributions from both transverse and longitudinal currents. Compared with the conventional space-charge solver, it introduces additional force terms that do not explicitly scale down with the relativistic factor \(\gamma\), because they arise from magnetic fields associated with the transverse current. This implies that these additional forces may not decrease as rapidly with beam energy as the conventional space-charge forces. In practice, however, these terms depend on the transverse velocity and therefore still decrease as the beam energy increases.

Using a converging Gaussian beam distribution, we derive a semi-analytical solution for the generalized space-charge forces. This analysis indicates that for the transverse remaining additional force to be significant compared with the conventional space-charge forces, \(\gamma\) must be on the order of \(10^{6}\). For the longitudinal remaining forces, the corresponding requirement is \(\gamma\sim 10^{4}\). For beams with an arbitrary density distribution, we present a numerical solver based on an integrated Green's function method.

As applications, we use the solver to study strongly converging electron beams in the interaction regions of FCC-ee and the ILC, where the electron beam is focused from sub-millimeter down to nanometer scales in a drift space. The results show that the additional energy spread induced by the generalized space-charge forces is small compared with the nominal energy spread assumed in the designs. This suggests that space-charge effects from highly converging beams do not pose a significant problem for these two colliders. This is primarily because the electron beam energies in FCC-ee and the ILC are already very high.

We further apply the generalized solver to electron-beam transport between the third and the fourth bending magnets in the second bunch-compressor chicane of the LCLS-II-HE design, where the horizontal RMS beam size decreases from millimeters to tens of microns. In this case, the generalized space-charge model does not lead to additional emittance or energy-spread growth relative to the conventional space-charge model
due to the small contribution from the remaining force. 

\section*{ACKNOWLEDGEMENTS}
 This work was supported by the U.S. 
Department of Energy (DOE), Office of Science, Office of 
High Energy Physics, under Contract No. DE-AC02- 
05CH11231. Computational resources were provided by 
the National Energy Research Scientific Computing Center 
(NERSC), which is supported by the DOE Office of 
Science under the same contract number.

\end{document}